\documentclass[]{aa} 
\usepackage{txfonts}
\usepackage{natbib}
\usepackage{graphicx, xspace}
\usepackage{multirow}

\usepackage{ amssymb }

\usepackage[switch]{lineno}

\newcommand{\kms} {\,km/s\xspace}
\newcommand{\vsini}{$v \sin i$\xspace}
\newcommand{\nm}{\,nm\xspace}
\newcommand{\sindex}{$\mathcal{S}$-index\xspace}
\newcommand{\Sindex}{$\mathcal{S}$-index\xspace}
\newcommand{\Ssymbol}{$\mathcal{S}$\xspace}
\newcommand{\num} {$\nu_{\rm max}$\xspace}
\newcommand{\logg} {$\log g$\xspace}
\newcommand{\dnu} {$\Delta\nu$\xspace}
\newcommand{\Kepler} {\textit{Kepler}\xspace}
\newcommand{\kepler} {\textit{Kepler}\xspace}
\newcommand{\Hermes} {\textsc{Hermes}\xspace}
\newcommand{\hermes} {\textsc{Hermes}\xspace}
\newcommand{\Mercator} {\textsc{Mercator}\xspace}

\newcommand{\twiny}{KIC\,3241581\xspace}
\newcommand{\muHz}{\,$\mu$Hz\xspace}
\newcommand{\dex}{\,dex\xspace}

\makeatletter

\newcommand{\Rmnum}[1]{\expandafter\@slowromancap\romannumeral #1@}
\makeatother

\usepackage[dvips]{color}

\newcommand{\new}[1]{\textbf{\blu{#1}}}
\newcommand{\add}[1]{\textbf{\blu{#1}}}
\newcommand{\blu}{\textcolor{blue} }
\renewcommand{\new}[1]{{#1}}
\renewcommand{\add}[1]{{#1}}
\renewcommand{\blu}[1]{{#1}}

\def\dnu{$\Delta\nu$\xspace}
\def\dn1{$\delta\nu_{01}$\xspace}
\def\dn2{$\delta\nu_{02}$\xspace}
\def\sun{\hbox{$_\odot$}\xspace}
\begin{document}
\title{The \Hermes Solar Atlas and the spectroscopic analysis of the seismic solar analogue \twiny\thanks{Based on observations made with the \Hermes spectrograph mounted on the 1.2\,m \Mercator Telescope at the Spanish Observatorio del Roque de los Muchachos of the Instituto de Astrof\'isica de Canarias.}}

\authorrunning{Beck et al.}
\titlerunning{The \Hermes Solar Atlas and the spectroscopic analysis of the seismic solar analogue \twiny.}

\author{P.\,G.~Beck\inst{1} \and C.~Allende\,Prieto\inst{2,3} \and T.~Van\,Reeth\inst{4} \and A.~Tkachenko\inst{4} \and G.~Raskin\inst{4} \and  \\ H.\,van\,Winckel\inst{4} \and {J.-D.~do\,Nascimento Jr.\inst{5,6}} \and {D.\,Salabert\inst{1}} \and {E. Corsaro\inst{1,7,8}} \and R.\,A.~Garc\'\i a\inst{1} } \mail{paul.beck@cea.fr}

\institute{Laboratoire AIM, CEA/DSM Ð CNRS - Univ. Paris Diderot Ð IRFU/SAp, Centre de Saclay, 91191 Gif-sur-Yvette Cedex, France. 
\and Instituto de Astrof\'{\i}sica de Canarias, E-38200 La Laguna, Tenerife, Spain. 
\and Departamento de Astrof\'{\i}sica, Universidad de La Laguna, E-38206 La Laguna, Tenerife, Spain. 
\and Instituut voor Sterrenkunde, KU Leuven, B-3001 Leuven, Belgium. 
\and Harvard-Smithsonian Center for Astrophysics, Cambridge, MA 02138, USA.
\and Departamento de F\'isica Te\'orica e Experimental, Universidade Federal do Rio 
\and Instituto de Astrof'sica de Canarias, 38205 La Laguna, Tenerife, Spain.
\and Universidad de La Laguna, Departamento de Astrof'sica, 38206 La Laguna, Tenerife, Spain.}

\abstract
{Solar-analog stars provide an excellent opportunity to study the Sun's evolution, i.e. the changes with time in stellar structure, activity, or rotation for solar-like stars.
The unparalleled photometric data from the NASA space telescope \Kepler allows us to study and characterise solar-like stars through asteroseismology.}
{We aim to spectroscopically investigate the fundamental parameter and chromospheric activity of solar analogues and twins, based on observations obtained with the \Hermes spectrograph and combine them with asteroseismology. Therefore, we need to build a solar atlas for the spectrograph, to provide accurate calibrations of the spectroscopically determined abundances of solar and late type stars observed with this instrument and thus perform differential spectral comparisons. }
{We acquire high-resolution and high signal-to-noise (S/N) spectroscopy to construct three solar reference spectra by observing the reflected light of Vesta and Victoria asteroids and the jovian moon Europa (100\,$\lesssim$\,S/N\,$\lesssim$\,450) with the \Hermes spectrograph. We then observe the \kepler solar analog \object{\twiny}(S/N$\sim$170). For this star, the fundamental spectral parameters are extracted in a differential analysis. Sufficient S/N in the near ultraviolet allows us to investigate the chromospheric magnetic activity in both objects.}
{We constructed three solar spectrum atlases from 385 to 900\nm obtained with the \Hermes spectrograph from observations of two bright asteroids and a jovian moon. A comparison between our solar spectra atlas to the Kurucz and HARPS solar spectrum shows an excellent agreement. \twiny was found to be a long-periodic binary system. The fundamental parameter for the stellar primary component are \add{T$_{\rm eff}$\,=\,5689\,$\pm$11K, $\log g$\,=\,4.385$\pm$0.005, $[$Fe/H$]$\,=\,+0.22$\pm$0.01}, being in agreement with the published global seismic values confirming its status of solar analogue. The chromospheric activity level is compatible to the solar magnetic activity observed during 2014 and 2015.}
{Our solar atlas is an essential tool for the analysis of solar-like stars and to characterise solar analogs and twins with \Hermes. The differential analysis, using the presented solar atlas from \Hermes observations allows us to obtain the fundamental parameters with a very high accuracy. KIC 3241581 is a metal rich solar analogue with a solar-like activity level in a binary system of unknown period.  }

\maketitle

\section{Introduction}

Solar-like stars and, in particular, solar analogues and twins provide an unique opportunity to understand our closest star, the Sun, in the context of stellar evolution. In the classical definition provided by \cite{CayreldeStrobel} and \cite{CayreldeStrobel1996} (hereafter CdS96), solar twins are stars that are spectroscopically and photometrically identical to the \object{Sun}, while solar analogues are stars within 10\% of the solar radius and mass, as well as metallicities within 0.3\,dex, but the definition allows for different ages.

The increasing number of known solar analogues allows us to study similar stars to the Sun but with differences in some of its properties such as stellar rotation, chemical composition or surface magnetic activity \cite[e.g.][]{Garcia2014b,Baumann2010,SchrijverZwaan2008}. Moreover, it is possible to place the Sun in its evolutionary context  as well as studying aspects of habitability  through direct observations of individual targets or following statistical approaches \citep[e.g.][]{Unterborn2015}. 

Recently, asteroseismic quantities have been included in the definition of solar analogs. However, the precise observation of solar-like oscillators is challenging, as the pulsations exhibit amplitudes of a few parts-per-million in photometry or at the level of meters-per-second in Doppler velocity. Therefore, ground-based observations are only possible with high accurate radial-velocity measurements \citep[e.g.][and references therein]{Aerts2010}. The advent of space telescopes, such as \textsc{CoRoT} \citep{Baglin2006} and \Kepler \citep{borucki2010}, allows us to study larger datasets of solar-like oscillators. Prime examples are the characterisation of some solar analogs such as 16\,Cyg\,A\&B from \kepler observations \new{\citep[e.g.][]{Metcalfe2012,doNascimento2014,Davies2015}} and \textsc{CoRoT}\,102684698 \citep{doNascimento2013}.

There are several convincing arguments to utilise asteroseismology to search for solar analogues and twins. The seismic key quantities of solar-like oscillations, the maximum oscillation power \num and the separation between consecutive radial modes \dnu correspond to the surface gravity and the mean sound speed and therefore allow to derive the mass and radius of a solar-like oscillator better than 10\% precision, as shown by large sample studies  for various evolutionary stages of single field stars \citep[e.g.][]{Chaplin2014,Kallinger2012}, cluster members \citep[e.g.][]{Hekker2011Cluster,Miglio2012,Corsaro2012}
and comparison with dynamical masses from double-lined binary systems \citep[e.g.][]{Frandsen2013}. A detailed analysis of the individual frequencies allows for an even more comprehensive study of the solar-like stars and solar analogues, including a detailed comparison of their stellar structure \citep[e.g.][]{Metcalfe2012,Metcalfe2014,Mathur2012, SilvaAguirre2015}.

To unfold its full potential, seismology needs to be combined with additional parameters from complimentary techniques. \cite{LebretonGoupil2014} as well as \cite{Chaplin2014} have demonstrated that very high accuracy is reached when combining seismology with the spectroscopically derived effective temperatures and abundances.
\new{For a subset of 90 solar-like oscillating stars, for which \cite{Bruntt2012} provided spectroscopic fundamental parameters, \cite{Chaplin2014} showed that using the actual chemical abundances for each star, instead of assuming a mean abundance of [Fe/H] = -0.2 for the Kepler field, that the precision of seismic parameters for each single star increases. }
\blu{The combination of seismology with a spectroscopically determined \new{parameters instead of} photometric calibrations improves the uncertainties by a factor of two, down to $\sim$5 and $\sim$2\,\% in mass and radius, respectively.} However, obtaining precise spectroscopic values for space targets is challenging, as many of those targets are rather faint (e.g. from the \kepler mission) and high-resolution spectroscopy is light demanding but requires high signal-to-noise ratios (S/N\,$>$\,100). In addition, detailed calibration is needed to achieve the highest precision possible. Using synthetic spectra for this purpose will limit the accuracy in the derived chemical composition of stars by typical $\sim$0.1\,dex, due to inaccuracies in atomic and molecular data that enter calculations of the model atmospheres. This restrain in the analysis of the stellar spectrum can be overcome by comparing it to the solar spectrum, obtained with the same instrument and setup in close temporal proximity. With this observing strategy, both spectra suffer from similar systematics, such as imperfections in the wavelength calibration and hence, the observations can be calibrated using the well-known parameters of the Sun.

To obtain the best spectroscopic calibrations for seismic studies and, in particular, for solar twins and analogs, efforts were spent on the construction of a calibrated solar reference spectrum, obtained with the \Hermes instrument. These reference spectra are made available for the public through this publication. 

Once this reference solar spectrum is built, we apply it to the study of the \kepler target \twiny. This star was  selected because it provides a close match to the solar global seismic parameters from the $\sim$ 500 \kepler stars analysed by \citet{Chaplin2014}. \twiny has also a surface averaged rotation period of around 26 days \citep{Garcia2014b} which implies that this star would be at around half of its main sequence evolution assuming general gyrochonology relations.

The analysis presented here has the purpose of a detailed spectroscopic study from high-resolution spectra obtained with the \hermes spectrograph, to test how well this seismically selected solar analogue resembles the CdS96 criterion for solar analogue abundances and to provide well constrained fundamental parameters for further theoretical modelling.

Finally, from spectroscopy we can also determine the level of stellar activity. The level of chromospheric activity as well as the duration of the activity cycles are age dependent. Typically, the chromospheric activity is measured from the emission of the core of the Ca\textsc{ii}\,H\&K lines in the near ultra violet as established by \citet[][and references therein]{Duncan1991} from the data from an extensive observing program at the Mount Wilson Observatory (MWO). Through a set of newly observed targets from the list of \cite{Duncan1991}, the \sindex for solar-like stars was calibrated for \hermes spectroscopy and applied to the solar analog \twiny.

\begin{figure*}[t!]
\centering
\includegraphics[width=0.999\textwidth,height=55 mm]{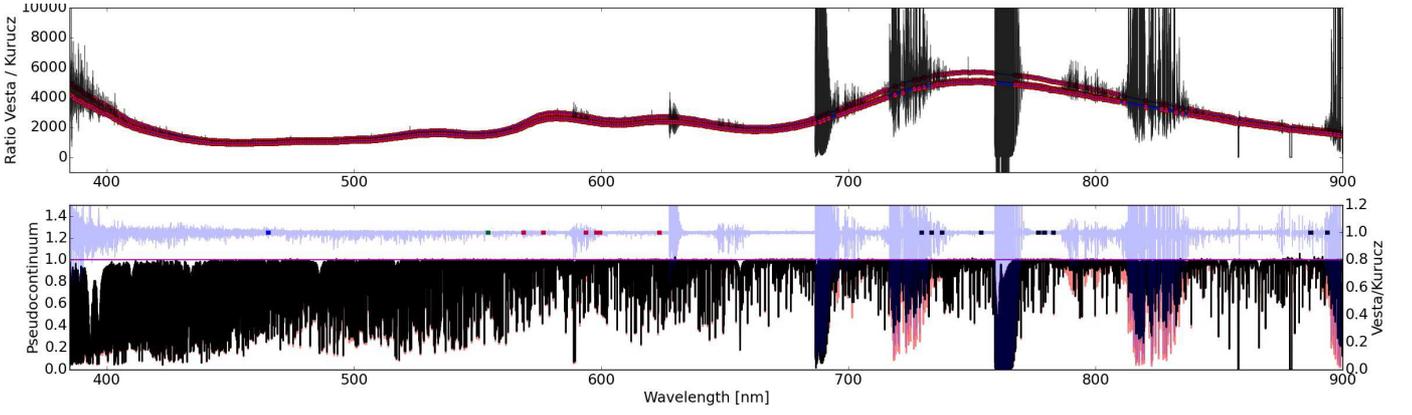}
\caption{Normalisation and compilation of the solar spectrum from observations of Vesta. The top panel shows the result of the division of the \Hermes spectra by the Kurucz solar atlas for each individual exposure. Red dots mark the supporting points for the cubic spline, depicted as blue line. The regions of large deviations from the fit originate from contamination through telluric lines. The bottom panel depicts the resulting median spectrum obtained for the Vesta observations (in black) and the \cite{Kurucz2005} spectrum (in red). The ratio between the spectra from Vesta and Kurucz are depicted as the blue line on the right hand y-axis. The regions used to determine the signal-to-noise ratio in Str\"omgren $b$, and $y$ as well as Johnson $R$ and $I$, are indicated in the bottom panel as blue, green, red and black dots, respectively.
\medskip\label{fig:Normalisation}}

\end{figure*}


\section{Observations \label{sec:Observations}}

\begin{table*}[t!]
\caption{\label{tab:observingLogVestaVictoria} Journal of observations of the solar spectra, obtained from the reflected light on the bright asteroids 4\,Vesta and 12\,Victoria}
\begin{center}
\begin{tabular}{ccccccr}
\hline\hline
Point &Date& Date  & Exp. time & Airmass & BVC &\multicolumn{1}{c}{RV}\\
source &[gregorian]& [HJD - 2456500.0] & [sec] & [] & [\kms] & \multicolumn{1}{c}{[\kms]}\smallskip \\
\hline
Vesta &2014, July 30& 369.374174 & 1200 & 1.48		& -29.171 & -10.710 $\pm$ 0.001\\ 
Vesta &2014, July 30& 369.388641 & 1200 &	1.59 		& -29.199 & -10.718 $\pm$ 0.001\\ 
Vesta &2014, July 30& 369.403109 & 1200 &	1.73		& -29.225 & -10.725 $\pm$ 0.001\smallskip\\ 
\hline
Victoria &2014, July 31& 	369.580444 & 1200 & 1.20	& +19.638 & 12.405 $\pm$ 0.001 \\ 
Victoria &2014, July 31& 	369.594914 & 1200 & 1.15	& +19.600 & 12.403 $\pm$ 0.002 \\ 
Victoria &2014, July 31& 	369.609384 & 1200 & 1.11 	& +19.561 & 12.405 $\pm$ 0.001\smallskip\\ 
\hline
\end{tabular}
\end{center}
\tablefoot{The heliocentric Julian date (HJD), the dimensionless airmass, as well as the barycentric velocity correction (BVC), calculated by the pipeline are given for he midpoint of the exposure time (Exp. time). The barycentric radial velocity (RV) and the corresponding uncertainty was determined from a weighted cross correlation as described in the text. All observations were taken in the night from July, 30 to July 31, 2014.\smallskip }
\medskip
\caption{\label{tab:observingLogEuropa} Observing sequences of the solar spectrum, obtained from the observations of the jovian moon Europa.}
\begin{center}
\begin{tabular}{ccccrcccr}
\hline\hline
Point &Date& Exp. time & N& &Date  & Airmass & BVC &\multicolumn{1}{c}{RV} \\
source &[gregorian]& [sec]&& & [HJD - 2456500.0]  & [] & [\kms] & \multicolumn{1}{c}{[\kms]}\smallskip \\
\hline
Europa & 2015, April 17 & 400 & 5 & start: & 630.442 	&1.15	&-28.713  &  -8.398\,$\pm$\,0.003 \\
& && & end: 						& 630.463	 	&1.21	&-28.747  &  -9.344\,$\pm$\,0.003\smallskip\\
\hline
Europa & 2015, April 17 & 200 & 19 & start: &630.467 	&1.24	&-28.758  &  -9.521\,$\pm$\,0.003\\
& && & end: 						& 630.519 	&1.71	&-28.855  &  -11.831\,$\pm$\,0.003\smallskip\\
\hline
Europa & 2015, April 19 & 200 & 32 & start: & 632.419 	&1.10	&-28.921  &  14.435\,$\pm$\,0.003\\
& && & end: 						& 632.514 	&1.72	&-29.103  &  18.029\,$\pm$\,0.003\smallskip\\
\hline
\end{tabular}
\end{center}
\tablefoot{The column $N$ reports the number of integrations taken in a sequence with a constant exposure time. The values of HJD, airmass and BVC (cf. Tab\,\ref{tab:observingLogVestaVictoria}) are reported at the beginning and end of each observing sequence.}
\end{table*}%
The spectroscopic observations of the reflected solar light and \twiny (V=10.35\,mag) which are presented in this paper were carried out with the \textit{High Efficiency and Resolution Mercator \'Echelle Spectrograph}  \citep[\Hermes, ][]{Raskin2011,RaskinPhD} mounted on the 1.2\,m \Mercator Telescope at the Observatorio del Roque de los Muchachos on La\,Palma, Canary Islands, Spain. The observations were obtained with the 2.5 arc seconds fibre for the high-resolution mode (HRF) with a resolving power of R$_{\rm \Hermes}$\,=\,$\lambda$/$\delta\lambda$$\,\simeq$\,85000 (whereby $\lambda$ and $\Delta\lambda$ are the wavelength and the width per wavelength bin, respectively), covering a wavelength range of 385\nm to 900\nm.  The reference wavelength calibration was obtained from the combined light from a hollow-cathode Thorium-Argon and a Neon arc lamps (hereafter referred to as \textit{ThArNe})

For the acquisition of a solar references spectrum, observations taken from a point source are better suited than those taken from the sky. The latter one suffers from scattering in the Earth atmosphere which alters the measured spectrum depending on the angle between the observed spot on the sky and the Sun \citep{Gray2000} and can modify the determined parameter by a few percent. Therefore, we followed the standard approach to obtain solar spectra from the light reflected from three different minor bodies in the solar system. For the asteroids 4\,Vesta~($\sim$7.4 mag) and 12\,Victoria~($\sim$9.4 mag) we obtained three spectra each in the night from July 30 to 31, 2014. This lead to an accumulated integration time of 1 hour per object. The journal of observations and the obtained radial velocities are given in Table\,\ref{tab:observingLogVestaVictoria}. In those nights spectra of \twiny were acquired as well.

Over a timebase of 426 days, 20 spectra of \twiny with a total integration time of 8.1\,hrs have been obtained with the same setup than for the solar light (including the nights in which the solar spectra were observed). More details about the monitoring are presented in Section\,\ref{sec:Twiny}.

To test the stability of the \hermes spectrograph and to acquire a very high S/N solar reference, we obtained another set of spectra from the jovian moon Europa with a total integration time of 3.4\,hours in additional runs in two nights between April 17 and 21, 2015. Due to the rather high radial velocity variation with respect to the asteroids, the total integration time of 4.3\,hours were divided into 56 individual exposures of 400 and 200 seconds, to minimise smearing effects. 
The details about the observing sequences are listed in Table\,\ref{tab:observingLogEuropa}.  
All observations of the solar light were carried out in moonless nights. The compilation and normalisation of the combined spectrum and the validation of the method are presented in Section\,\ref{sec:MasterSpectrum}.

The obtained raw spectra were reduced and wavelength calibrated with the current version of the \Hermes data reduction pipeline \citep[Version 5, ][]{Raskin2011}. During the reduction of the spectra with the instrument-specific reduction pipeline, the spectra are rebinned to a constant resolving power of R=85000. The combined one-dimensional spectrum has no wavelength gaps, except for two small sections between \'echelle orders in the infrared between 857.7 and 858.1\nm, as well as 878.6 and 879.5\nm. 

For all individual spectra the radial velocities were obtained through a weighted cross-correlation of the wavelength range between 478 and 653\nm of each spectrum with a  G2 template \citep{Raskin2011,Raskin2014}. From the analysis of long time spectroscopic series of red giants, it was recently shown by \cite{Beck2014a,Beck2015a}, that with this approach \Hermes is capable of an excellent precision, allowing us to study solar-like oscillations with amplitudes of a few meters-per-second.


\begin{figure*}[t!]
\centering
\includegraphics[angle=90,width=0.95\textwidth,height=0.93\textheight]{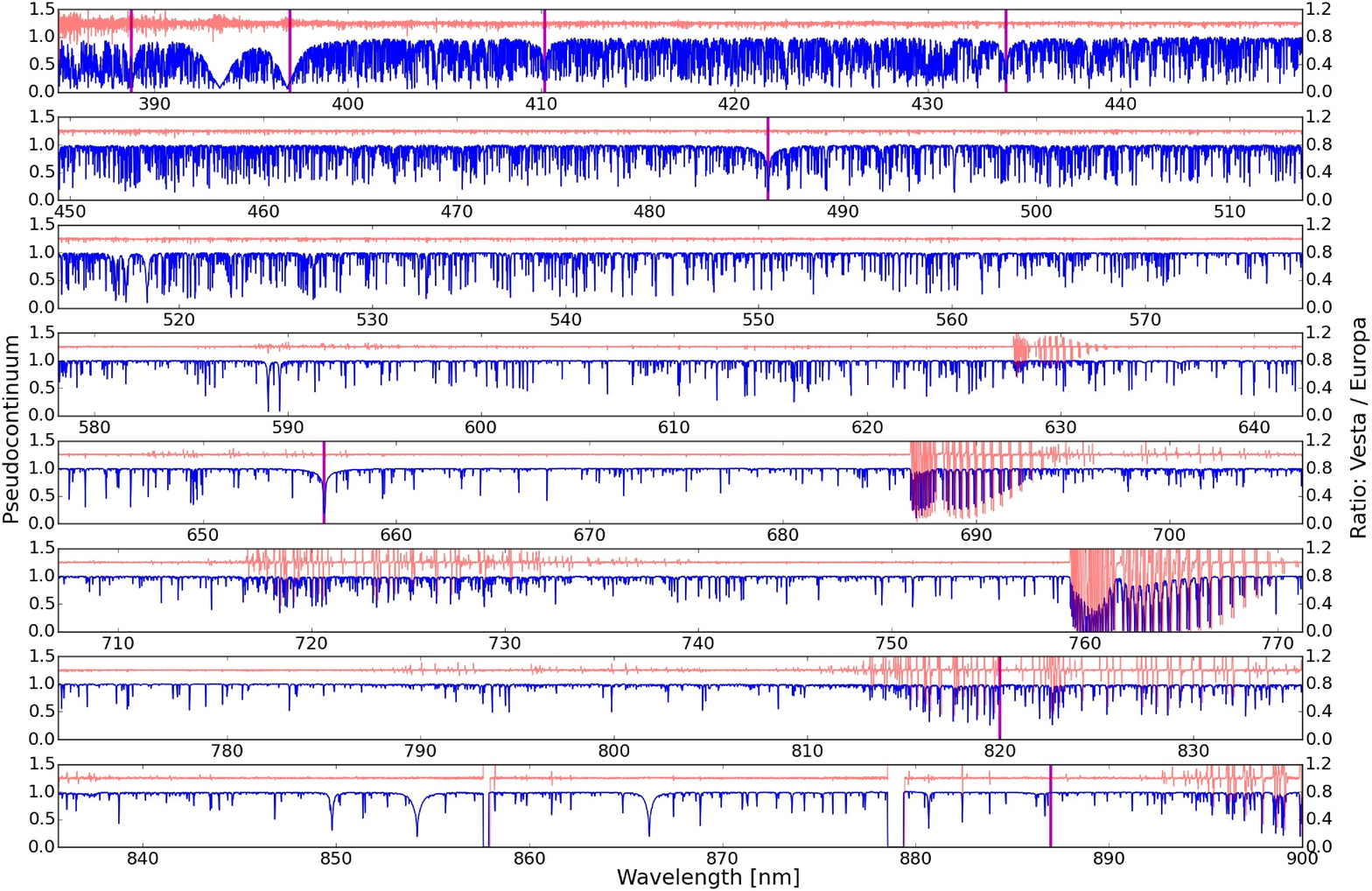}
\caption{The combined spectra of Europa plotted (blue). The flux ratio per wavelength bin between the two median spectra of Europa and Vesta is plotted in red. Regions of large deviations reveal strong contamination through telluric lines. The position of the Hydrogen lines from the Balmer and Paschen series are marked with vertical magenta bars. }
\label{fig:medianSpectra}
\end{figure*}

\begin{figure*}[t!]
\centering
\includegraphics[width=\hsize]{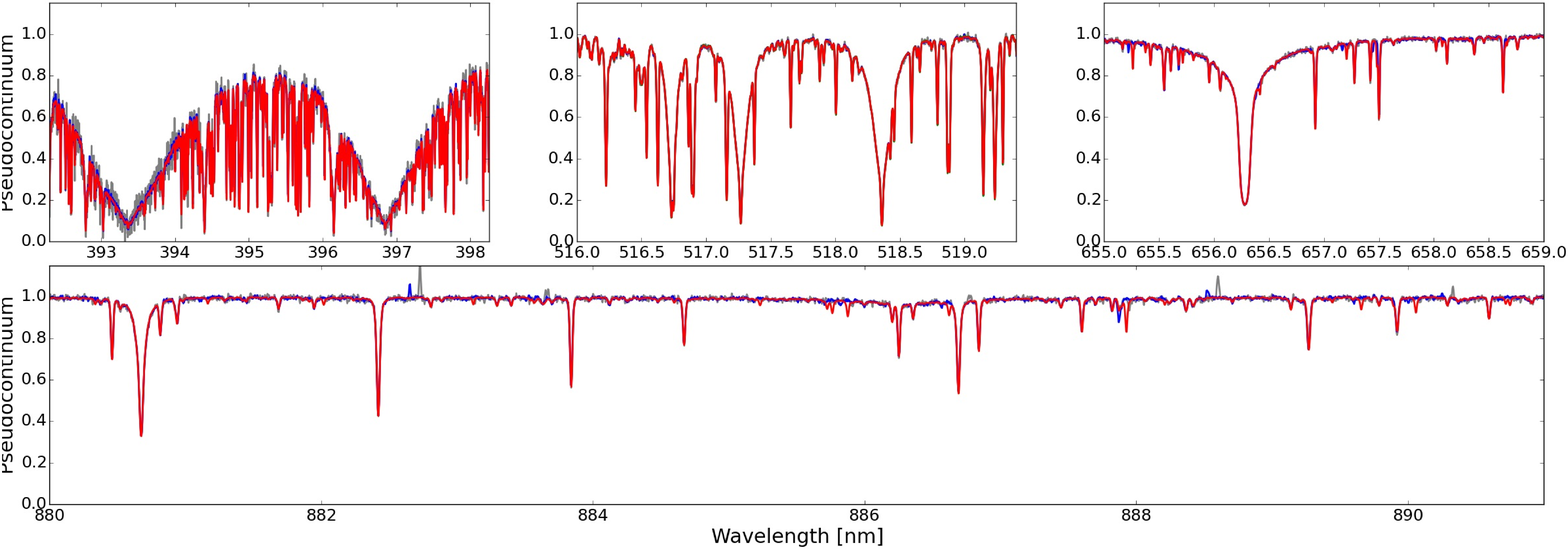}
\caption{Comparison of the \Hermes spectra (blue and grey lines) with the smoothed Kurucz spectrum (red line) around the Ca H\&K lines (top left panel), the Mg-triplet (top centre panel) and the H$_\alpha$ line (top right panel). The bottom panel shows the region in the infrared around the Paschen line at 887\nm. The differences between the three spectra around H$_\alpha$ and the Paschen line originate from the contamination through telluric lines. In the infrared, also several weak telluric emission lines are seen. \label{fig:spectrumSnapshots}}
\end{figure*}


\section{\label{sec:MasterSpectrum} The solar spectrum}

The availability and accuracy of the atomic and molecular data that enters calculations of model atmospheres and synthetic spectra is limited. In addition, such modelling usually involves a fair number of approximations such as hydrostatic  and radiative equilibrium, and local thermodynamical equilibrium. All these factors limit the accuracy in the derived chemical compositions of stars, typically to $\sim$0.1 dex.
However, differential studies involving stars that are similar regarding atmospheric parameters can achieve a much higher precision. As an example, differential studies for solar analogues or twins \citep[e.g.][]{Ramirez2009,Melendez2014,Nissen2015} are routinely delivering relative abundances good to $\sim$0.01-0.02 dex.

The Sun is the only star for which we know the atmospheric parameters (effective temperature and surface gravity) to very high accuracy: about 3 K for Teff and 0.0001 dex for logg \citep{Stix1991}. Hence,  a precise differential analysis  of a solar-like star relative to the Sun, can be immediately translated into an accurate analysis. As we show in the tests below, we can measure with a precision of  a few degrees the effective  temperature of a star like the Sun relative to the Sun. Likewise, we can constrain the surface gravity of such star to an excellent precision. Therefore, we can determine those values to high accuracy, thanks to our knowledge of the absolute values of those quantities for the Sun.

To be able to perform such studies also for stellar targets, observed with the \Hermes spectrograph, we also aimed to obtain solar reference spectra for this instrument and understand its characteristic. It also shall be tested, if the simultaneous acquisition of reference spectra is needed in the same night or if the reference spectra can be used over a longer time frame.

\subsection{Compilation of the solar atlas \label{sec:compilation}}
All reduced \Hermes object spectra were corrected for the motion of the earth and the barycentric radial velocities measured from cross correlation are given in Table\,\ref{tab:observingLogVestaVictoria}. 
To remove the trend remaining in the reduced one-dimensional spectra before compiling the combined spectrum, we divided the individual spectrum by the solar spectrum using the solar atlas by \cite{Kurucz2005}, appropriately smoothed as a reference. For more details on the reference spectrum, we refer to Section\,\ref{sec:literatureAtlas} in this paper.
A cubic spline was fit through $\sim$500 supporting points, that were equally spaced to map the long periodic variation in the ratio between the stellar and the reference spectra (Fig\,\ref{fig:Normalisation}, top panel). Each data point represents the median flux value in a window of 0.5\nm. In the wavelength ranges where telluric lines are present, the spectra deviate strongly. For these regions, supporting points were placed manually to fit the continuum regions. Finally, the observed spectrum was divided by the cubic spline. By adopting this technique, we force the observed spectrum to follow the envelope of the pseudo continuum of the reference spectrum. This approach also corrects for the decrease of the intensity in the ultra-violet due to unresolved lines.
 The combined solar spectrum from each \new{individual} source (Fig.\,\ref{fig:Normalisation}, bottom panel) was compiled by calculating the median flux from the individual spectra for each wavelength bin. \new{We note that only spectra from the same source are merged.}
 
The resulting spectrum from the reflected light from Europa is presented in Figure\,\ref{fig:medianSpectra}.
The location of telluric lines is visible through the strong outliers in the flux ratio between spectra of Vesta and Victoria (red line). Figure\,\ref{fig:spectrumSnapshots} illustrates the quality of the spectra for several regions, important for the spectroscopic analysis. In all observations, the width of the telluric lines was determined to be $\sim2.5$\kms, from the sigma of the gaussian fit to the average telluric line profile. Also several telluric lines are found in emission. However, the intensity of this atmospheric \textit{night glow} \citep[e.g.][]{Hilliard1966} hardly impacted the observed spectra.

\begin{table}[t!]
\caption{Signal-to-noise values in specific spectral pass bands. \label{tab:SN}}
\begin{center}
\begin{tabular}{ccccc}
\hline\hline
Wavelength & N & 4\,Vesta & 12\,Victoria & Europa \\
\hline
Acquisition year: & & 2014 & 2014 & 2015 \\
\hline
Str\"omgren $b$ 	&4	& 210\,$\pm$20 	& 110\,$\pm$10		& 470\,$\pm$200 \\
Str\"omgren $y$ &4	& 280\,$\pm$30 	& 130\,$\pm$10	& 440\,$\pm$130 \\
Johnson $R$ 		&9	& 270\,$\pm$40	& 150\,$\pm$20	& 460\,$\pm$150 \\
Johnson $I$ 		&9	& 250\,$\pm$40 	& 130\,$\pm$20	& 430\,$\pm$170 \\
\hline
\end{tabular}
\tablefoot{The reported S/N represents the mean S/N of $N$ continuum regions in the spectrum. The standard deviation from those subsets is given as uncertainty. The passband gives the location in which the regions without stellar or telluric lines, that were used to estimate the S/N for the compiled solar median spectrum from the observations of Vesta and Victoria.}
\end{center}
\label{default}
\end{table}%

Each median spectrum such as shown for Europa in Figure\,\ref{fig:medianSpectra}, corresponds to a total integration time of 3600 seconds, leading to a general signal-to-noise ratio  ranging between 110 and 470 for the different spectra.  The S/N was determined from regions, (ideally) without stellar or telluric absorption lines,
\begin{eqnarray}
S/N &=& \frac{1.}{std(F/F_{mean})}, \label{eq:sn}
\end{eqnarray} 
whereby $std(F/F_{mean})$ is the standard deviation of the flux $F$ in the selected region, renormalised  through the mean flux $F_{mean}$.

To find regions suited for the computation of the S/N, we compared the median spectra from \Hermes observations to the solar atlas by Kurucz as well as a synthetic spectrum of the Sun. This comparison showed that there are hardly any regions that are free from absorption lines, forcing us to adopt regions with minimal contamination too. Therefore, the reported S/N is likely to be biased towards lower values through the contamination from small lines. Table\,\ref{tab:SN} lists the S/N found for both median spectra, determined from such small snippets of continuum with respect to their corresponding classical photometric pass bands. As it was calculated from several small sections of the spectrum, a mean S/N value is reported, as well as the spread among the values.


\subsection{Testing the consistency of \Hermes solar spectra from different sources and times}

For a better understanding of the systematics contained in our observations, we compare the distributions of the difference of the flux per wavelength bin between the solar spectra obtained from \Hermes and the solar atlases in literature. 

Possible differences between the solar spectrum obtained from Europa, Vesta and Victoria could be due to instrumental and atmospheric variations, or due to different characteristics of the asteroids, such as their albedo. 
The histograms of the differences of Europa and Victoria with respect to Vesta are shown in the left panel of Figure\,\ref{fig:histograms}. Comparing the normalised, accumulated histograms (right-hand y-axis) with the accumulated theoretical Gaussian function shows that the distributions of the differences resemble a Gaussian distributions well. The peak of the Gaussian distribution (Tab.\,\ref{tab:distributionParameters}) is centred as expected at 0 within the resolution of the histogram ($\sim$0.08\%).

Small differences in the raw data originating from observations at different air masses (Tab.\,\ref{tab:observingLogVestaVictoria} and \ref{tab:observingLogEuropa}) are removed by the normalisation to the pseudocontinuum. Since the observations of the three objects were separated in time, instrumental setup and the number of sun spots could in principle vary. No changes were made in the instrumental setup between the two observing runs. The international sunspot number, reported by the National Geophysical Data Center, NOAA\footnote{http://www.ngdc.noaa.gov/stp/solar/solardataservices.html.} for 2014 July, 30 and 31 were 88 and 95, respectively. For the observations of Europa the sunspot number was in a similar range (2015, April 17/18: 85 and 100, respectively; April 20/21: 93). Those are rather small changes and will not strongly affect the result of the two spectra.

The two main systematic effects contributing to the scatter are the presence of telluric lines and the different levels of S/N. The width of the distribution of the histograms in Figure\,\ref{fig:histograms} is dominated by the lower S/N in the spectrum of Victoria and also by the decreasing S/N towards the blue part of both spectra (cf. Fig.\,\ref{fig:medianSpectra}). Due to their position, depending on the stellar radial and barycentric velocity, telluric lines lead  to a large number of outliers.  Therefore, the histograms were calculated for the full spectral range (left panel, cyan distribution) and excluding the regions that are most contaminated by telluric lines (red). Both distributions are centred on 0 within the histogram resolution.
 
Finally, we show the differences between Europa and Vesta in the left panel of Figure\,\ref{fig:histograms}. For simplicity of the diagram only the spectra cleaned from heavy telluric contamination are compared. The ratio between the two spectra is also shown over the full range in red on the right hand side of Figure\,\ref{fig:medianSpectra}. Although nearly separated by 3/4 of a year, the distribution of the differences is well centred on 0, indicating that the quality of the spectrograph is stable and only depends on the S/N obtained in the final spectra.

\subsection{Comparing solar \hermes spectra to solar atlases in the literature \label{sec:literatureAtlas}}
To compare our three solar median spectra from \Hermes with published high fidelity solar atlases in the literature, we performed similar differential analyses with the revised FTS Kitt Peak Solar Flux Atlas of \cite{Kurucz2005} and the laser calibrated solar atlas by \cite{Molaro2013}, obtained from observations of the moon with the HARPS spectrograph. 

\begin{figure*}[t!]
\includegraphics[width=1\textwidth]{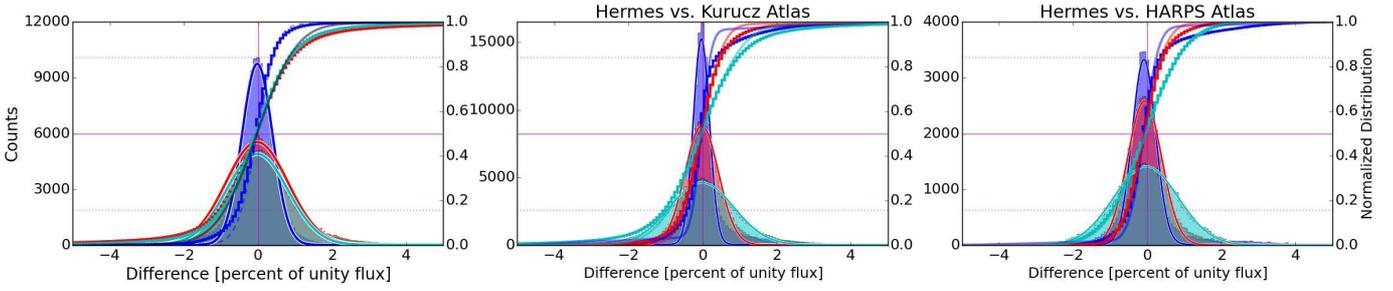}
\caption{Differential analysis of the spectra. The left panel shows the comparison between \Hermes spectra of Europa, Vesta and Victoria. The cyan histogram represent the distribution over the full spectral range, while the red histogram were calculated with the strongest telluric contribution excluded. The blue histogram indicates the distribution of the differences between the spectra of Vesta and Europa.
The centre panel compares the two spectra from \Hermes observations to the degraded Kurucz spectrum. In the right panel the \Hermes spectra are compared to the \textsc{Harps} spectrum. In the centre and right panels, the blue, red, and cyan distributions depict the comparison of the reference spectra with the observations from Europa, Vesta and Victoria, respectively. The same cleaning for telluric lines has been applied. 
The thin solid vertical and horizontal lines mark the theoretical centre of the distribution, while the dashed lines mark the borders of the 1$\sigma$ (68.27\%) interval.  Accumulated and normalised histograms are shown on the right axis. The thick grey line depicts the shape of the accumulated theoretical Gaussian distribution.
\label{fig:histograms}}
\end{figure*}

The very high S/N spectrum of \cite{Kurucz2005} was compiled from 50 Fourier Transform Spectroscopy (FTS) scans and extends beyond the blue and red borders of our spectra. The resolution of the spectrum ranges from R =400\,000 to 500\,000 and therefore had to be degraded to approximately fit the resolution R$_{\rm \Hermes}$. 
The atlas of \cite{Molaro2013} is covering a wavelength range between 475 and 586\nm with a gap around 532\nm. 
The original resolution of R=115\,000 was recalculated to R$_{\rm \Hermes}$. Since the published version of the spectrum is not normalised, we had to split the original file at 532\nm and normalise both \'echelle orders the same way as described for our solar spectra by utilising the Kurucz spectrum as a template for the pseudocontinuum.

\begin{figure}[t!]
\begin{center}
\centering
\includegraphics[width=0.49\textwidth]{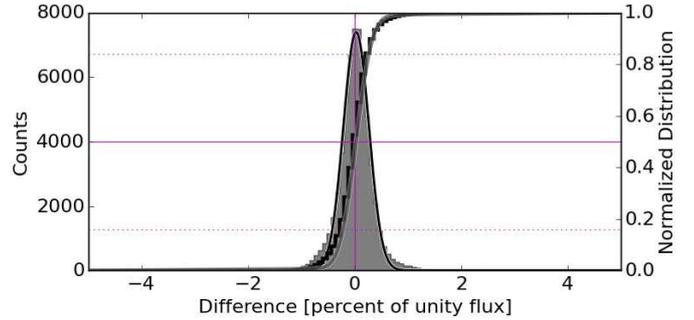}
\caption{Distribution of the differences between the Kurucz and the \textsc{Harps} literature spectra.}
\label{fig:literatureDistribution}
\end{center}
\end{figure}

The distributions comparing the median \Hermes spectra to the Kurucz and the renormalised \textsc{Harps} atlases are shown in the centre and right panel of Figure\,\ref{fig:histograms}.
All distributions are centred at zero (Table\,\ref{tab:distributionParameters}).  
The differences in the width of the distributions originates from the difference in S/N among the three \Hermes spectra (Tab.\,\ref{tab:SN}). Finally, Figure\,\ref{fig:literatureDistribution} compares the distribution of the differences between the two literature atlases, smoothed to the spectral resolution of \Hermes. Here the peak is centred on 0.03$\pm$0.24.

\subsection{Data products}
From the comparison of the \Hermes spectra among each other, as well as the differential analysis with two published solar atlases, we conclude that the \Hermes spectra are not significantly different. The level of the signal-to-noise found in both spectra is substantially higher than what is typically achieved in stellar spectra. Also the stability of the instrument over a longer time base was confirmed, indicating that the acquisition of solar calibration in the same night  as the data is not needed to achieve high accuracy in the analysis.

We also find that our normalisation procedure is working satisfactorily and delivers a pseudocontinuum that resembles the observations of \cite{Kurucz2005}. The only main source of the scatter and outliers are the differing levels of S/N and the telluric lines respectively.


\begin{table}[t!]
\caption{Centre and width of the Gaussian distributions of the residuals.}
\tabcolsep=2.8pt
\centering
\begin{tabular}{lcccc}
\hline\hline
& Kurucz & \textsc{Harps} & \Hermes(full) & \Hermes(cleaned)\\
& [percent] & [percent] & [percent] & [percent]\\
\hline
Europa & -0.03\,$\pm$0.25&-0.08\,$\pm$0.30 &\\
&&&$-$& -0.02\,$\pm$0.43\\
Vesta &  -0.02\,$\pm$0.45 & -0.07\,$\pm$0.46 & \\
&&&-0.03\,$\pm$0.84 & -0.02\,$\pm$0.83\\
Victoria &  -0.02\,$\pm$0.86  & -0.03\,$\pm$0.87\\
Lit.\,Ref. &  \multicolumn{2}{c}{+0.03\,$\pm$0.24} \\

\hline
\end{tabular}
\tablefoot{The table reports the centre and width of the distribution of the residuals between different sources. The columns '\hermes (full)' and '\hermes (cleaned)' report on the residuals between two obtained with \Hermes. The row 'Lit. Ref.' gives the distribution of the differences between the two reference atlases, used in this work (cf. Fig.\,\ref{fig:literatureDistribution}).}
\label{tab:distributionParameters}
\end{table}%

To provide further users of the \Hermes instrument with the possibility to calibrate their observations through the solar spectrum, the three compiled solar atlases obtained from Vesta and Victoria in 2014 and from Europa in 2015 and presented in this work are made available in the online material of the paper. The full data package, including also raw and wavelength calibration data, will be available in an electronic archive on the webpage of the \Mercator telescope\footnote{http://www.mercator.iac.es/ \textit{(webpage under construction)}}: 

\begin{itemize}
\item The normalised median spectra for the solar spectrum (385 - 900\nm), obtained from Vesta, Victoria, and Europa as described in the paper are provided. 
\item The individual, unnormalised integrations (377 - 900\nm) are available as one dimensional FITS-files. Those spectra are uncorrected for the radial and barycentric velocity. Several ThArNe spectra are available to determine the instrumental broadening of \Hermes.
\item The synthetic solar spectrum, calculated for infinite spectral resolution. Also the line list to identify the chemical elements for the absorption lines is given.
\item A selection of figures for teaching and public outreach will be presented on this page following the example of \cite{Beck2015SW}.
\end{itemize}

\begin{table*}[t!]
\caption{Fundamental parameters of the Sun and \twiny.}
\centering
\tabcolsep=15pt
\begin{tabular}{l|llll|l}
\hline\hline
\multirow{2}{*}{Parameter}   & \multicolumn{1}{c}{Solar reference}&  \multicolumn{3}{c}{\Hermes Solar spectrum from source} &  \multicolumn{1}{c}{Solar analogue} \\
&   \multicolumn{1}{c}{\cite{Kurucz2005} } &\multicolumn{1}{c}{Europa} & \multicolumn{1}{c}{Vesta}& \multicolumn{1}{c}{Victoria}&\multicolumn{1}{c}{\twiny}\\
\hline
T$_{\rm eff}$	 [K]	 		&5767$\pm$9		& 5788$\pm$3		&5774$\pm$3 	&5769$\pm$4	& 5689$\pm$3	 \\
$\log g$		 [dex]		&4.452$\pm$0.005 	& 4.453$\pm$0.006	& 4.432$\pm$0.006	&4.426$\pm$0.005 	& 4.385$\pm$0.005	 \\
$[$Fe/H]		 [dex]		& 0.000$\pm$0.002	&0.002$\pm$0.002	& -0.001$\pm$0.002	&0.000$\pm$0.002 & 0.220$\pm$0.002	 \\ 
\hline
\end{tabular}
\tablefoot{
The effective temperature T$_{\rm eff}$, projected surface velocity \vsini, the surface gravity $\log g$ and the iron abundance [Fe/H]. The solar reference  values for those parameters of the Sun, derived from \cite{Kurucz2005}. The deviation to the literature values of a given parameter, derived from the spectra atlases of Europa, Vesta and Victoria as well as the uncertainty of the measurement are given. In the last column the fundamental parameter of the seismic solar analogue \twiny are given. The error ranges reflect the internal uncertainties of the method \add{and are discussed in the text}.}
\label{tab:fundamentalParameters}
\end{table*}%

In addition to the three already existing data sets of Vesta, Victoria and Europa, new solar atlases from ongoing monitoring of the mentioned targets as well as additional minor bodies (such as Ceres) will be added to the database to provide reliable reference spectra for different epochs. Such growing sample of objects provides a selection of prominent celestial calibration targets and at least one of the objects will be visibly throughout the year, allowing to compare new observations to the archived data.

We aim for frequent reobservation of those targets during our yearly campaigns to further test the stability of the instrument. For the intermediate future, there are plans by the \hermes-team to further maximise the flux efficiency of the fibre link between the telescope and the \Hermes spectrograph and to optimise the wavelength calibration over the full wavelength range through instrumental updates. New calibrated spectral references atlases will be provided for the revised instrumental setup. Such archive will provide observers at or users of the \Hermes data archive with a well calibrated solar spectrum, taken in close temporal proximity and identical instrumental set up in numerous years. A consistent sets of standard-spectra allows also to compare the progress in the optimisation of the instrument.

The choice of standard spectral reference spectra is not limited to the solar case only.  The optimal results are found if the star from which the 'golden standard' spectrum originates from is as close as possible in mass, temperature and radius to the target star. Such templates for a differential analysis would also allow an easier extraction of the activity tracers in late type stars.

We therefore aim to extend the archive beyond main sequence, solar analogue stars, such as to $\gamma$\,Doradus stars \citep[e.g.][]{Tkachenko2013b} or stars in various phases of the red giant phase. Prime examples would be Arcturus (K0III) or the well described bright red giant in the wide binary system of \object{$\theta^1$\,Tau} (=vb77\,Tau, K0III), which which is also a member of the very well studied Hyades cluster \citep{Perryman1998,deBruijne2001}. This red giant has been characterised in detail through spectroscopic disentangling techniques from \Hermes observations and seismic diagnostics from radial velocities time series \citep[][respectively]{Beck2015c,Beck2015a}. 
Such reference spectra will also ease the process of normalisation or could serve as instrument specific cross correlation templates for such stars.

\section{The Sun as a reference \label{sec:SunAsStar}}
To characterise the uncertainties that arise from using solar spectra as a reference for calibrating the spectroscopic libraries or in a differential analysis, we tested what the uncertainties are from using these spectra in a sun-as-star experiment. The values of the Sun-as-star experiment (Table\,\ref{tab:fundamentalParameters}) were based on our solar spectra from Europa, Vesta and Victoria in the wavelength range from 509 to 522 nm, which includes the Mg\,\textsc{i} Triplet at 517 nm. In addition, we also experimented with the \cite{Kurucz2005} solar atlas, smoothed to the resolving power of the HERMES observations (R=85000).

In our analysis approach we compare the obtained spectra to a grid of synthetic spectra, with the 3 free parameters: T$_{\rm eff}$, \logg and [Fe/H], with a fixed micro=2.0 km/s, ignoring macro turbulence. The analysis is carried out using {\tt FERRE} \citep{AllendePrieto2006}\footnote{FERRE is available from hebe.as.utexas.edu/ferre}. In the first step, we interpolated in the  grid of synthetic spectra to obtain a model appropriate for the solar parameters (T$_{\rm eff}$=5777\,K, \logg= 4.437 and [Fe/H]=0.00). We then modify our grid of models, multiplying all the model fluxes by the ratio of the observed and interpolated fluxes for the Sun. The resulting grid of models is therefore 'forced' to match
the solar observations, the median of \hermes fluxes for Europa, Vesta and Victoria, at the solar parameters.
The optimal solutions were found with FERRE using the UOBYQA algorithm \citep{Powell2000} and performing Bezier quadratic interpolation on the model grid. 
\add{Our estimated formal uncertainties for both
the Sun and KIC 3241581 compare favorably with those reported by
Nissen (2015) for a set of solar analogs and a differential analysis
relative to the Sun based on equivalent widths (6 K in T$_{\rm eff}$, 0.01 in logg}
The results for the fundamental parameters and the [Fe/H] abundance from the solar spectra in this Sun-as-Star experiment are listed in Table\,\ref{tab:fundamentalParameters}, whereby the uncertainties correspond to 1-$\sigma$ random errors derived by inverting the curvature matrix. Naturally, we find a good agreement with the literature values and the uncertainties are dominated by the differing values of S/N of each combined solar spectrum. 

This approach allows us to carry out a differential analysis of the KIC\,3241581 relative to the Sun. In addition to the internal uncertainties provided by FERRE, we computed 'empirically' our uncertainties from the dispersion found for the parameters of KIC\,3241581 and the Sun, for which we have one and four spectra (Vesta, Victoria, Europa, and the Kurucz et al. solar atlas) available, respectively. It is noteworthy that the intrinsic uncertainties and those we find empirically are fairly similar, \add{and the values are within 11\,K for T$_{\rm eff}$, 0.002\,dex for [Fe/H], and about 0.01\,dex for logg to the solar value. We therefore adopt these values as more conservative uncertanties for the parameters.} These tiny uncertainties demonstrate the potential of a differential analysis relative to the Sun.
The small scatter derived for the multiple solar spectra supports the small error bars for the derived parameters. In Section\,\ref{sec:Twiny} we will demonstrate that since the atmospheric parameters of the Sun extremely precisely known, that we are able to derived  accurate parameters for KIC\,3241581.

\section{The calibration of the \Sindex for \hermes spectroscopy of solar-like stars}
\begin{table}[t!]
\caption{Comparison of the \Sindex for solar-like MWO-calibrator stars \label{tab:stellarSindexCalibration}}
\centering
\tabcolsep=2.5pt
\begin{tabular}{llcccc}
\hline\hline
\multicolumn{1}{c}{Object} & \multicolumn{1}{c}{SpecTyp} 	 & \multicolumn{2}{c}{\sindex} & \multicolumn{1}{c}{Spec} & \multicolumn{1}{c}{Residuals}  \\ 
& &  \multicolumn{1}{c}{MWO}&\multicolumn{1}{c}{\hermes} &&  \multicolumn{1}{c}{$\mathcal{S}$-$\alpha\cdot\mathcal{S_{\rm H}}$}\\\hline
\object{HD66171}	& G2V	& 0.188$\pm$0.012	& 0.0083	& 1			& 	0.003 \\
\object{HD88737} &F9V 		& 0.233$\pm$0.006	& 0.0102	& 1			& 	0.002 \\
\object{HD114710} &F9.5V 	& 0.200$\pm$0.005	& 0.0088$\pm$0.0001	& 2		& 	0.002\\
\object{HD115383} &G0V 		& 0.313$\pm$0.014	& 0.0136$\pm$0.0001	& 3		& 	0.000\\
\object{HD115043} & G2V 	& 0.317$\pm$0.027 	& 0.0138$\pm$0.0001& 2		& 	0.000\\
\object{HD120136} &F6IV 	& 0.188$\pm$0.005	& 0.0082$\pm$0.0003	&6		& 	0.001\\
\object{HD146233}$^\star$ & G2Va 	& 0.174$\pm$0.010	& 0.0078$\pm$0.0029	& 12	& 	0.006\\ 
\object{HD190771} & G2V		& 0.334$\pm$0.028	& 0.0148	& 1 		& 	0.006\\
\hline

\end{tabular}
\tablefoot{The identifier, spectral type and the total number of observations of the calibrator stars in the \hermes archive are given. The \sindex was calculated from the combined median \Hermes data spectrum, following the procedure described in Section\,\ref{sec:MasterSpectrum}. The values reported for the Mount Wilson Observations (MWO) was calculated from the mean and standard deviation of all tabulated observations by \cite{Duncan1991}. The error bars for the \hermes observations are calculated from the dispersion between the \Ssymbol calculated for individual spectra. The residual between MWO and \hermes are are given. $\star$: also known as 18\,Sco.}
\bigskip
\caption{The chromospheric activity index \Ssymbol for the Sun from observations of the reflected light.}
\centering
\tabcolsep=3pt

\begin{tabular}{llccc}
\hline\hline
\multicolumn{1}{c}{Object} & \multicolumn{1}{c}{Date} & \multicolumn{2}{c}{\sindex} & Sun Spot \\ 
& \multicolumn{1}{c}{[gregorian]} & \Ssymbol$_{\hermes}^{\rm original}$ & \Ssymbol$_{\hermes}^{\rm MWO}$&Number\smallskip\\ \hline

\new{Moon} 	& \new{2009, August 04}		& \new{0.0077}		& \new{0.18}	&\new{0} \\
Vesta 	& 2014, July 30 		& 0.0083 		& 0.19	& 88-100\\
Victoria	& 2014, July 31 		& 0.0080		& 0.18	& 88-100\\
Europa	& 2015, April 17\&20 	& 0.0083		& 0.19 	& 85-93\smallskip\\\hline
\twiny 	& 2014-2015 			& 0.0077		& 0.18	& $-$\smallskip\\
\hline
\end{tabular}
\tablefoot{Object used as source of reflected sun light and the Date of observations. The instrumental, MWO-scaled \sindex are given and compared to the international sun spot number.}
\label{tab:solarSindex}
\end{table}%

To quantify the activity level in late-type stars, we also aimed to determine the \Sindex, an activity indicator based on the chromospheric emission in the core of the Ca\textsc{ii}\,H\&K absorption line in the near ultra violet. Those lines are also included in \Hermes spectra.

To monitor the chromospheric activity of stars and to compare it to the large database of observations in the literature, we utilised the formalism of the \sindex of the Mount Wilson Observatory \citep[MWO,][and references therein]{Duncan1991} that quantifies the level of chromospheric activity in the Ca\textsc{ii}\,H\,\&\,K lines,
\begin{eqnarray}
\mathcal{S}_{\rm \Hermes} = \frac{N_H+N_K}{N_R+N_V}.
\label{eq:sIndexHermes}\end{eqnarray}
In this formalism, $N_H$ and $N_K$ are the fluxes in the centres of the H and K lines, weighted by a triangular filter with a FWHM of 0.109\nm centred on 396.847 and 393.368\nm, respectively. This flux is compared with the continuum flux the two spectral segments $N_R$ and $N_V$ of 2\nm, centred on 390.107 and 400.107\nm, respectively.

Originally designed for a photon-counting spectrograph using photomultipliers, the fluxes corresponded to counts in the photometric passbands after the classical corrections (dead-time correction and sky correction). In the modern one-dimensional spectra, the surface under the pseudocontinuum replaces the counts in the photometric filters. The dimensionless indicator is dependent on the spectral resolution as well as the colour index and the metallicity of the stellar object and needs to be scaled through the multiplicative constant $\alpha$ onto the original value system for each instrument by using calibrator stars from the list of \cite{Duncan1991},
\begin{eqnarray}
\mathcal{S}^{\rm MWO}_{\rm Hermes} = \alpha_{\rm solar} \cdot \mathcal{S}_{\rm Hermes}, 
\label{eq:sIndexCalibration}\end{eqnarray}
Therefore, we obtained data for a a set of 8 calibrator stars from the original catalogue of MWO in an observing run in March and July 2015, as well as by searching the \hermes data archive.
For these standard stars, the mean value for \Ssymbol from the literature, as well as the measured \sindex from \hermes spectra is listed in Table\,\ref{tab:stellarSindexCalibration}. The errors for MWO reflect the dispersion among measurements, reported in the catalogue.  From fitting a linear regression to the these pairs of values, we find that for solar-like stars, $\alpha_{\rm solar}$=23$\pm$2. The main source of uncertainty in this comparison is the varying activity level in those solar-like stars, modulated by activity cycles whereby we compare observations likely at different phases. Therefore a scatter is expected. The found linear fit gives a good result.  The \new{fit derived has residuals} better than 2\%.  %
\begin{figure}[t!]
\centering
\includegraphics[width=0.45\textwidth]{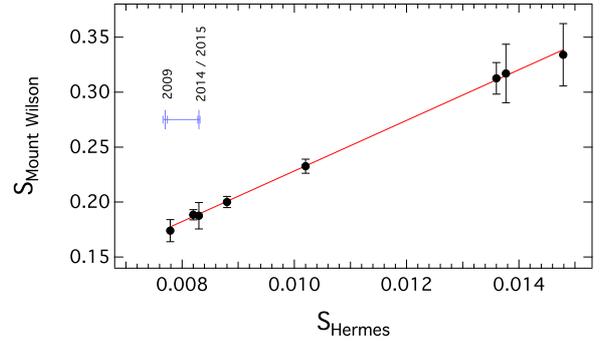}
\caption{Calibration of the \sindex by comparing the measured \sindex from Hermes with the average tabulated value from MWO-observations (cf. Table\,\ref{tab:stellarSindexCalibration}). For multiple observations, the vertical error bars reflect the scatter of those stars in the MWO-catalogue, otherwise, the reported uncertainty was adopted. The red line depicts the linear correlation. For comparison, the instrumental \Ssymbol-value of the Sun, \new{measured in minimum phase of cycle 23 in August 2009 and in maximum phase of cycle 24} April 2015, on the horizontal axis is shown, \new{reflecting the variation }. The scatter of \Ssymbol within the time series of our observations of Europa is shown with the \new{blue} symbol in the upper left part of the diagram. }
\label{fig:sIndexCalibration}
\end{figure}

The \Sindex found from \new{our} observations of the Sun via Vesta, Victoria and Europa are reported in Table\,\ref{tab:solarSindex}. These observations were conducted during the maximum of solar activity \new{of the solar cycle 24}. To test the intrinsic scatter in high-S/N spectra, we calculated the dispersion of \Ssymbol in the 56 spectra from observations of Europa in April 2015 (Table\,\ref{tab:observingLogEuropa}), and found a typical scatter of $\pm$0.0002. For comparison of the Sun to the set of calibration stars, the measured solar-activity level (Table\,\ref{tab:solarSindex}) is marked on the horizontal axis of Figure\,\ref{fig:sIndexCalibration}. \new{Naturally the \sindex of the Sun follows the solar activity cycle. In the archive of \Hermes, we found unpublished observations of the solar spectrum, using the moon as reflecting source from August, 4 2009. These observations were conducted during the recent minimum of solar activity. From those 12 spectra we measure a mean value and dispersion of $\mathcal{S}$\,=\,0.077$\pm$0.004, respectively. Figure\,\ref{fig:sIndexCalibration} compares the value form the solar minimum in 2009 to the \Sindex of the recent (moderate) maximum in 2014/15, revealing how tiny the variation of \Ssymbol is among the solar cycle.}

\section{Analysis of the seismic solar analogue \twiny \label{sec:Twiny}}
Seismic modelling as well as evolutionary tracks are very sensitive to the chemical composition. Therefore, it is interesting, to investigate, how well a seismic solar analogue, selected from seismological indications, such as \twiny, does fulfil the third criterion by CdS96, constraining the maximum deviation of the metal abundances to less than 0.3\dex. 
We obtained high-resolution spectroscopy with the \Hermes spectrograph. 
To test for binarity or variation in the tracers of chromospheric activity, we monitored this star with 20 observations spread over $\sim$1.5 years.

\begin{figure}[t!]
\centering
\includegraphics[width=\columnwidth]{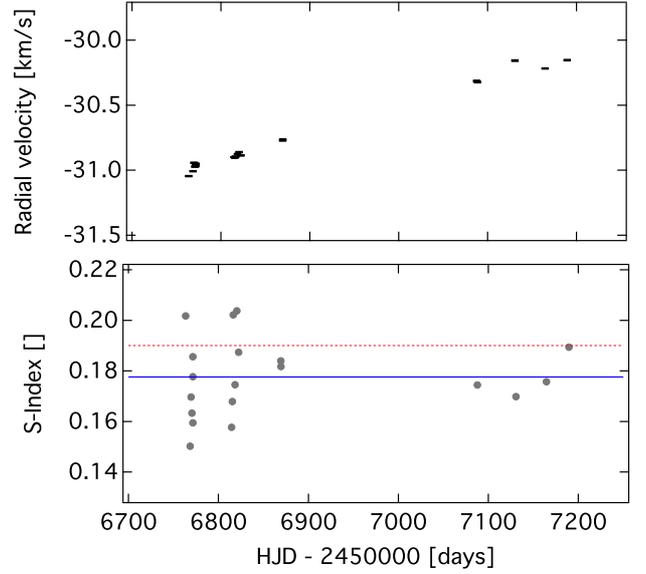}
\caption{Spectroscopic monitoring of \twiny. top panel: radial velocities of \twiny. 
bottom panel: The variation of the chromospheric activity indicator, $\mathcal{S_{\rm Hermes}^{\rm MWO}}$ of \twiny. The red line indicates the \Sindex of the Sun from April 2015. \label{fig:twinyRV} }
\end{figure}

The radial velocities obtained from the individual spectra, depicted in Figure\,\ref{fig:twinyRV} (top panel) and listed in Table\,\ref{tab:twinyRV} reveal that \twiny is a binary system with a period substantially longer than 1.5 year. The measured difference of currently more than {900\,m/s} in radial velocity excludes a component with a substellar mass. A visual inspection of the spectra  as well as the average line profiles from cross correlation of the stellar spectrum with a G2 mask (cf. Section\,\ref{sec:Observations}) did not show a spectroscopic signature of the secondary component in the individual composite spectra. Also no trace of a secondary is found in the power-spectral density. In the context of this paper, we refer to the primary of this system when using the KIC identifier. The photometric dilution due to blending \citep{Miglio2014,Johnston2015} could be the cause for a lower S/N in the oscillation power spectrum. Once the orbit and its elements are resolved, this object is an interesting candidate for the application of spectral disentangling techniques \citep[e.g. FDBinary,][]{Ilijic2004}.

\begin{table}[t!]
\caption{Journal of observations and radial velocities of \twiny. }
\centering
\tabcolsep=10pt
\begin{tabular}{ccc}
\hline\hline
\multicolumn{1}{c}{HJD-2456500} &	\multicolumn{1}{c}{RV [km/s]} & \Ssymbol\\
\hline
263.61191 & 		-31.049\,$\pm$0.003	 	& 0.202\\
268.65827 & 		-31.033\,$\pm$0.003 	& 0.151\\
269.64710 & 		-30.944\,$\pm$0.003 	& 0.170\\
270.65539 & 		-30.987\,$\pm$0.002 	& 0.163\\
271.59569 & 		-30.979\,$\pm$0.003 	& 0.178\\
271.64209 & 		-30.960\,$\pm$0.004 	& 0.160\\
271.66548 & 		-30.964\,$\pm$0.004 	& 0.186\\
314.71648 & 		-30.905\,$\pm$0.004 	& 0.158\\
315.51440 & 		-30.913\,$\pm$0.002 	& 0.168\\
316.50856 &		-30.904\,$\pm$0.002 	& 0.202\\
318.51228 & 		-30.892\,$\pm$0.003 	& 0.174\\
320.49913 & 		-30.876\,$\pm$0.003 	& 0.206\\
322.51981 & 		-30.905\,$\pm$0.002 	& 0.188\\ 
369.42749 & 		-30.783\,$\pm$0.003 	& 0.184\\
369.55871 & 		-30.771\,$\pm$0.003 	& 0.182\\ \hline
587.75066 & 		-30.311\,$\pm$0.002 	& 0.175\\
588.75505 & 		-30.334\,$\pm$0.009 	& 0.080\\
630.68656 &		-30.201\,$\pm$0.006		& 0.168\\
664.64064 &		-30.235\,$\pm$0.005		& 0.175\\ 
689.60817 & 		 -30.154\,$\pm$0.005	& 0.190\\ \hline
\hline
\end{tabular}
\tablefoot{The heliocentric Julian date (HJD) is given for he midpoint of the exposure. The barycentric radial velocity (RV) and the corresponding uncertainty were determined from a weighted cross correlation as described in the text. }
\label{tab:twinyRV}
\end{table}%

\begin{figure*}[t!]
\begin{center}
\includegraphics[width=\textwidth]{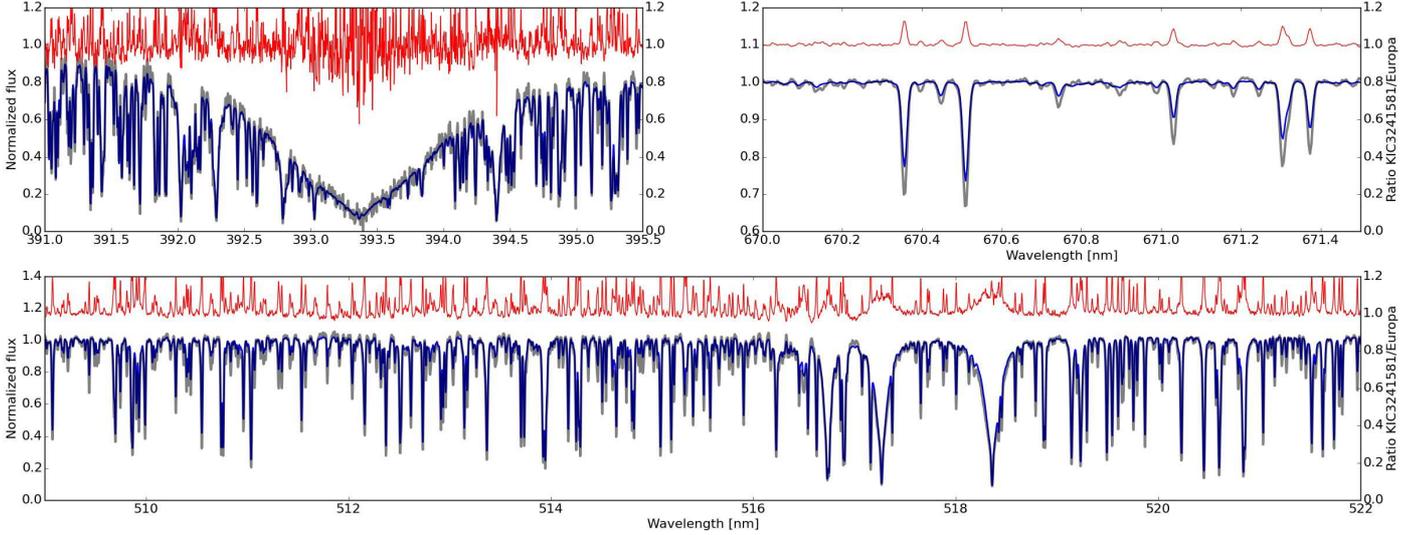}
\caption{Direct comparison of the spectra of the solar analogue \twiny (grey line) through the spectrum of Europa (blue). The value of the spectrum, normalised to the pseudo continuum is given on the left y-axis. The red curve on top of the two spectra depicts the ratio between the flux per bin of \twiny and Vesta (right y-axis). The Ca\,H line, the Mg triplet and the lithium multiplet at 670.8\,nm are shown in the left, middle and right panel, respectively.}
\label{fig:diffTwiny} 
\end{center}
\end{figure*}

Combining the 20 individual spectra from monitoring following the same procedure as described for the solar reference spectra, we obtain a median spectrum with a total integration time of 8.1\,hours and a S/N of $\sim$170 in Str\"omgren $y$. In Figure\,\ref{fig:diffTwiny}, several important segments of the final spectrum of \twiny are shown and compared to the solar spectrum, obtained from Europa.

The fundamental spectroscopic parameters of the solar analogue were determined from the spectral range between 509 to 522\,nm, which surrounds the Mg triplet. Following the same analysis approach as described for the Sun-as-a-Star in Section\,\ref{sec:SunAsStar} \add{using the combined average of the three (reflected) solar spectra}, the fundamental parameters of \twiny were determined to be T$_{\rm eff}$=5689$\pm$3, 
$\log$g=4.385$\pm$0.005, and [Fe/H]=0.220$\pm$0.002. Yet, the retrieved [Fe/H]-abundance is still well within the range described by CdS96.

From the scaling relations of \cite{Chaplin2011}, the seismic values \new{reported by Garcia et al. (in prep)},  \num=2751\muHz, \dnu=122.9$\pm$1.6\muHz, and the effective temperature \new{derived in this work} reveal a star of 1.03$\pm$0.1\,M\sun, 1.08$\pm$0.1\,R\sun and $\log g$=4.39$\pm$0.01, which is well within the 10\%-range in mass and radius for solar analogues, defined by CdS96. \new{The analysis of Garcia et al. (in prep) has improved the seismic analysis of \cite{Chaplin2014} by an improved data processing \citep{Garcia2014} as well as the extraction of individual oscillation modes instead of using global seismic pipelines.} Using the surface temperature and the radius determined from spectroscopy and asteroseismology, respectively, we find a stellar luminosity of 1.096\,L\sun. The surface rotation rate remains stable with a period of P$_{\rm rot}$$\simeq$26\,days \citep{Garcia2014b} and the value of photometric activity index, Sph=177.2\,ppm are close to the solar values of P$^\odot_{\rm rot}\simeq$24.47\,days and Sph$_\odot$=166.2\,ppm (Salabert et al., in prep).


The fundamental parameter which is best constrained through global seismology is the surface gravity, $\log g_\star$ since it is directly proportional to \num of the star and can be constrained better than 2\%. It only further depends on the square root of the temperature ratio, $T_{\rm eff}/T^\odot_{\rm eff}$. Also the surface gravity is in agreement from the findings of spectroscopy and seismology.

The average \sindex from the combined stellar spectrum of \twiny, which is $\mathcal{S}$\,=\,0.177, is found slightly below the values found for the  $\mathcal{S}$-value from our observations of the Sun (Table\,\ref{tab:solarSindex} and Figure\,\ref{fig:twinyRV}, bottom panel). Given the spread of the variations of the solar activity, \twiny shows a typical solar activity level. Such a stellar activity as well as the period of the surface rotation rate are indicating that the star is a main sequence star \citep[e.g.][]{Skumanich1972,Saar1999}. Recently, for solar-like stars observed with the \kepler satellite, \cite{Karoff2013} correlated ages derived from theoretical, seismic models with the Ca\,H\&K excess flux and provided a scaling relation for both quantities and confirms the result of previous studies.

The lithium abundance of KIC\,3241581 (Figure\,\ref{fig:diffTwiny}, top right panel) was derived from the Li\textsc{i} resonance transition at 670.7\nm. A synthetic spectrum was fitted to the \hermes spectrum, for the set of atmospheric parameters presented at the Table\,\ref{tab:fundamentalParameters}.  T$_{\rm eff}$ = 5689\,K, $\log g$ = 4.385\,dex, [Fe/H] = +0.220, and $\zeta$ = 1.21\kms.   Model atmospheres were interpolated in the Kurucz grid \citep{Kurucz1993} and the synthetic spectra were calculated with the MOOG routine \citep{Sneden73}. The synthesis of Fe\textsc{i} lines in the $\lambda$670.7 vicinity provides a low projected rotational velocity \vsini. 
For the spectroscopic atmospheric parameters\footnote{In this work, we use the standard definitions: $[X/Y]$=$\log$(N$_X$/N$_Y$)-$\log$(N$_X$/N$_Y$)$_\odot$, and $A_X$=$\log$(N$_X$/N$_H$)+12, where $N_X$ is the number density of element X in the stellar photosphere.}, we derived  \add{A$_{\rm Li}$ }=$\log$N(Li) = 0.03 \add{dex. 
The value A$_{\rm Li}$$\sim$ 0.03 is at the same scale and typical for a solar-analuges with a slightly higher mass than the solar value \citep{doNascimento2009}.   
For a typical 1 solar mass at the end of the PMS, the typical value of Lithium abundance is  around A$_{\rm Li}$$\sim$3.0. }
\add{The derived} abundance value is \add{therefore} well in agreement with  \twiny being on the main sequence.

 
\section{Discussion \& Conclusions}

To achieve the best accuracy possible for the spectroscopic analysis of solar-like stars and solar analogues and twins, we obtained solar reference spectra from light reflected from three minor bodies in the solar system, 4\,Vesta, 12\,Victoria and Europa. The data products were validated through the direct comparison with two other solar atlases from the literature, as well on the basis of the derived fundamental parameters. An excellent agreement with found in both approaches. From the observation of solar-like stars, we calibrated the \Sindex, a tracer for the chromospheric magnetic activity level as described by \citet[][and references therein]{Duncan1991} for the \hermes spectroscopy.
Following the procedure of the compilation of the median spectrum, which is outlined in the paper, new solar reference catalogues will be produced to have recent solar spectra in code proximity to the ongoing monitoring in our solar analogue observing program.

The solar analogue \twiny was identified from its close match of the global seismic quantities, \num and \dnu, and the surface rotation rate derived from photometric time series collected with the \Kepler space telescope. From scaling relations, the star is found to be slightly more massive than the Sun, but being well within the boundaries of the definition of \citet[][and references therein]{CayreldeStrobel1996}.  

The star has been monitored since more than 1.5 years, in the high-resolution observing mode of the \Hermes spectrograph, revealing that \twiny is the primary of a long periodic binary system. The spectroscopic fundamental parameters as well as the metal abundance were derived from a differential analysis. Using the solar spectrum, obtained from observations of the jovian moon Europa as a template for the differential analysis, we determined the fundamental parameters to be T$_{\rm eff}$\,=\,5689$\pm$3\,K, $\log g$\,=\,4.385$\pm$0.005, $[$Fe/H$]$\,=\,+0.220$\pm$0.002 and found that the star is indeed within the interval metallicity interval given by the classical definition for solar analogues by CdS96. Further analysis showed the chromospheric  \Ssymbol activity index of \twiny as well as the lithium abundance are good agreement with being a main sequence star.

In addition to the high-precision space photometry from space mission such as \kepler and K2, ground-based spectroscopy at intermediate size telescopes such as \Mercator or the currently built \textsc{Song} telescope network \citep{Song2011}
are providing crucial and independent information to understand the nature of the stars, such as abundances or the variability of the level of chromospheric activity. Having a high-precision calibration tool, such as the high S/N solar atlas available 
will help to better characterise such targets and derive accurate abundances as the input parameters for stellar models, 
as we have shown for \twiny.

\begin{acknowledgements}
The observations are based on spectroscopy made with the \Mercator Telescope, operated on the island of La Palma by the Flemish Community, at the Spanish Observatorio del Roque de los Muchachos of the Instituto de Astrof'sica de Canarias. PGB acknowledges the ANR (Agence Nationale de la Recherche, France) program IDEE (n¡ ANR-12-BS05-0008) "Interaction Des Etoiles et des ExoplanetesÓ. TVR acknowledges financial support from the Fund for Scientific Research of Flanders (FWO), Belgium, under grant agreement G.0B69.13. {AT is a Postdoctoral Fellow of the Fund for Scientific Research (FWO), Flanders, Belgium.}
E.C. is funded by the European CommunityÕs Seventh
Framework Programme (FP7/2007/2013) under grant agreement No 312844
(SPACEINN). The research leading to these results has received funding from the European Community's Seventh Framework Programme ([FP7/2007-2013]) under grant agreement No. 312844 (SPACEINN) and under grant agreement No. 269194 (IRSES/ASK). PGB also received funding from the CNES grants at CEA. The authors thank Dr. Antonio Garcia for providing us with the archived solar spectra from 2009.
We thank the anonymous referee for constructive criticism which helped to improve the paper.

\end{acknowledgements}

\bibliographystyle{aa}
\bibliography{/Users/pbeck/publications/inProgress/bibliography.bib}

\end{document}